\documentstyle[11pt,amssymb]{article}

\def\be{\begin{equation}}
\def\ee{\end{equation}}

\title{{\hfill\small\tt Theor.\,Math.\,Phys.\,123,\,671-672\,(2000)}\\
{\hfill}\\
A remark on the matrix Airy function}

\author{A.M. Perelomov\,
\footnote{\,On leave of absence from Institute for Theoretical and
Experimental Physics, 117259 Moscow, Russia. Current e-mail address:
perelomo@dftuz.unizar.es}
\\{\small\em Max-Planck-Institut f\"ur Mathematik,}\\
{\small\em Vivatsgasse 7, 53111 Bonn, Germany}}

\date{}
\begin{document}
\maketitle

\begin{abstract}
An integral representation for matrix Airy function is presented.
\end{abstract}
\medskip
\rightline{\sl In memory of M.V. Saveliev}
\bigskip

\setcounter{equation}{0} \noindent The well-known Witten
conjecture [1], that the logarithm of the partition function of
general one-matrix model can be considered as the generating
function for intersection numbers on the moduli space of algebraic
curves, was proved by Kontsevich [2,3], who reduced this
problem to the computation of the matrix Airy function (see 
[4--6], where many aspects of this problem were discussed). The
matrix Airy function is defined by an integral representation. We
give another integral representation for this function.

Following [3], we define the matrix Airy function on the space
${\cal H}_N$ of Hermitian $N \times N$ matrices by the integral
\be A(X)=\int \mbox{exp}\,\left\{ i\left( \frac13\,\mbox{tr}(Y^3)-
\mbox{tr}(XY)\right) \right\} dY \ee 
over the space ${\cal H}_N$,
where $dY$ is a $U(N)$-invariant measure on this space, $X,Y\in
{\cal H}_N$. We also note that the function $A\,(X)$ satisfies the
equation \be \Delta \,A\,(X) + (\mbox{tr}\,X)\, A(X) =0, \ee where
$\Delta$ is the Laplacian.

We recall that the classical Airy function
\be
\mbox{Ai}\,(x) = \int _{-\infty} ^{+\infty} \exp \left( i( y^3/3 -xy)\right)
dy \ee
is the unique solution (up to a scalar factor)  of the differential equation
\be
\mbox{Ai}^{''}(x) + x\,\mbox{Ai}(x) = 0 \ee
that is bounded on real axes. It follows from Eq.(1) that the dependence of
$A(X)$ on the variables $\tilde X$ and $\xi = N^{-1}\,\mbox {tr}\,X$, where 
$X=\tilde X+\xi I$ and $\mbox{tr}\,\tilde X=0$, can be separated. Indeed we have
\[
X = \tilde X + \xi I,\qquad Y = \tilde Y + \eta I,\qquad \mbox{tr}\,
\tilde X = \mbox{tr}\, \tilde Y = 0, \]
\[
\mbox{tr}\,(Y^3)=\mbox{tr}\,(\tilde Y^3)+3\eta \,\mbox{tr}(\tilde Y^2)+
N\eta ^3, \]
\[
\mbox{tr}\,(XY) =\mbox{tr}\,(\tilde X\tilde Y) + N\xi \eta .\]

{\bf Theorem 1.} {\em The integral representation}
\begin{eqnarray}
A(X)&=&A(\tilde X,\xi )=A(Q,\xi )\\
&=& CN^{-1/3}\int \exp \left\{
\frac{i}3\,\mbox{tr}\,(P^3)\right\}\mbox{Ai}\left( N^{-1/3}(N\xi
-\mbox{tr}(P^2))\right) \Phi (Q|P)\,d\mu (P) \nonumber
\end{eqnarray}
{\em is valid for  matrix Airy function} (1), {\em where} $Q$ {\em  and}
$P$ {\em are diagonal matrices}, $\tilde X=UQU^{-1}$, $\tilde Y=U_1P
U_1^{-1}$, $U$ {\em and} $U_1$ {\em are unitary matrices},$\Phi (Q|P)$ 
{\em is the zonal spherical function}
\be \Phi (P|Q)=\frac{\mbox{det}\,\{\exp(i(q_jp_k))\}}{V(P)\,V(Q)}\,. \ee
\be
Q=\mbox{diag}\,(q_1,\ldots ,q_N),\qquad P=\mbox{diag}\,(p_1,\ldots ,p_N),
\qquad d\mu (P)=(V(P))^2\,dP, \ee
$C=\mbox{const}$, {\em and} $V(P)$ {\em is the Vandermonde determinant.}
\be
V(P)=\prod _{j<k}(p_j-p_k)=\mbox{det}\,(p_j^{k-1}). \ee

{\bf Proof.} This follows immediately from the Harish-Chandra formula.

{\bf Theorem 2} [7]. {\em If} $\Phi $ {\em is a conjugacy
invariant function on the space Hermitian} $N\times N$ {\em
matrices, then }
\begin{eqnarray}
&&\int \Phi (Y)\,\exp(-\,i\,\mbox{tr}\,(QY))\,dY \nonumber \\
& =&(-2\pi i)^{N(N-1)/2}(V(Q))
^{-1}\int \Phi (P)\,\exp(-\,i\,\mbox{tr}(QP))\,(V(P))^2dP 
\end{eqnarray}
{\em for any diagonal Hermitian matrix} 
$Q$, {\em where} $Y=UPU^{-1}$
{\em and last integral is taken over the space of diagonal
Hermitian matrices} $P$.

We thus obtain an integral representation of the matrix Airy
function such that the dependence on the variable $\xi $ is
contained only in the standard one-dimensional Airy function. This
representation may be useful for investigation of properties of
the matrix Airy function.

For the case $N=2$, we have $\mbox{tr}(P^3)=0$, and integral
representation (5) takes the simpler form \be A(X)\equiv A(\xi
,r)=\int \exp \left\{ \frac{2i}3\,(\eta ^3+3\eta p^2- 3\xi \eta
)\right\} \Phi (r|p)\,p^2\,dp\,d\eta , \ee where \be \Phi
(r|p)=\frac{\sin pr}{pr} \ee is the zonal spherical function for
the matrix space with $N=2$. The integration over $\eta $ gives
\be A(\xi ,r)=C\int \mbox{Ai}\left( \left(
\frac32\right)^{1/3}(\xi -p^2)\right) \Phi (r|p)\,p^2dp,\qquad
C=2\left( \frac32\right) ^{1/3}\,. \ee If we take the integral
over $p$ in (10), we obtain another integral representation \be
A(\xi ,r)=C_1 \int \exp \left\{ \frac{2i}3\,(\tau ^3-3\xi \tau
)\right\} G(r,\tau )\,d\tau ,\qquad C_1=\mbox{const}, \ee where
\be G(r,\tau )=\tau ^{-3/2}\,\exp \{ i\tau ^{-1}r^2\} \ee is the
Green function of the Schr\"odinger equation for free motion in
three-dimensional space.

{\bf Acknowledgments.} This note was written during a stay at
 Max-Planck-Institute f\"ur Mathematik, Bonn. It is pleasure 
 to thank the staff of the institute for their hospitality.

\end{document}